# Connecting Defects and Amorphization in UiO-66 and MIL-140 Metal-organic Frameworks: A Combined Experimental and Computational Study

Authors: Thomas D. Bennett[1]*, Tanya K. Todorova[2], Emma F. Baxter[1], David. G. Reid[3], Christel Gervais[4], Bart Bueken[5], B. Van de Voorde[5], Dirk De Vos[5], David A. Keen[6] and Caroline Mellot-Draznieks[2]*

1 Department of Materials Science and Metallurgy, University of Cambridge, 27 Charles Babbage Road, Cambridge, CB3 0FS, UK.

2 Laboratoire de Chimie des Processus Biologiques, UMR 8229 CNRS, UPMC Univ Paris 06, Collège de France, 11 MarcelinBerthelot, 75231 Paris Cedex 05, France

3 Department of Chemistry, University of Cambridge, Lensfield Road, Cambridge, CB2 1EW, UK

4 Sorbonne Universités, UPMC Univ Paris 06, CNRS, Collège de France, UMR 7574, Chimie de la Matière Condensée de Paris, Paris, France

5 Centre for Surface Chemistry and Catalysis, Leuven Chem&Tech, KULeuven, Celestijnenlaan 200F p.o. box 2461, 3001 Heverlee, Belgium

6 ISIS Facility, Rutherford Appleton Laboratory, Harwell Oxford, Didcot, Oxon, OX11 0QX, UK.

# Abstract

The mechanism and products of the structural collapse of the metal-organic frameworks (MOFs) **UiO-66**, **MIL-140B** and **MIL-140C** upon ball-milling are investigated through solid state $^{13}$C NMR and pair distribution function (PDF) studies, finding amorphization to proceed by the breaking of a fraction of metal-ligand bonding in each case. The amorphous products contain inorganic-organic bonding motifs reminiscent of the crystalline phases. Whilst the inorganic $Zr_6O_4(OH)_4$ clusters of **UiO-66** remain intact upon structural collapse, the ZrO backbone of the **MIL-140** frameworks undergoes substantial distortion. Density functional theory calculations have been performed to investigate defective models of **MIL-140B** and show, through comparison of calculated and experimental $^{13}$C NMR spectra, that amorphization and defects in the materials are linked.

# Introduction

Crystalline metal-organic frameworks (MOFs) continue to be of interest to the scientific community due to their high surface areas and related potential for gas sorption, separations, drug delivery and catalysis.[1,2] Comparatively little attention is however focused on their amorphous counterparts, i.e. those which do not exhibit long range crystallographic order, yet still consist of three-dimensional arrays of connected inorganic nodes and organic ligands.[3-5] The use of pressure, temperature or shear stress to induce transitions between crystalline and amorphous states is of particular intrigue,[6-8] due to possible uses in reversible sorption, conductive and multiferroic applications.[9] Whilst computational or experimental characterization of purely inorganic[10] or organic[11] amorphous frameworks is known, structural insight into amorphous MOFs (*a*MOFs) is largely limited to those which adopt similar network topologies to the zeolite family.[12]

Mechanochemistry, or ball-milling is increasingly utilized to synthesize crystalline MOFs in relatively large quantities with minimal solvent use.[13] Curiously, MOFs are highly susceptible to collapse using the same treatment.[14] Whilst this can be mitigated by pore-filling prior to treatment, it remains problematic given the use of ball-milling as a common post-processing method to increase external surface area.[15, 16] This propensity for collapse has previously been linked to the low shear moduli of the family,[17] although recent reports have emerged that appear to contradict this argument.[18, 19]

The zirconium-based MOF **UiO-66** $[Zr_6O_4(OH)_4(O_2C-C_6H_4-CO_2)_6]$[20] crystallizes in the space group $Fm\bar{3}m$, and consists of $Zr_6O_4(OH)_4$ octahedra connected together by benzene-1,4-dicarboxylate (**bdc**) linkers in three dimensions.[21] The inorganic cluster has a twelve-fold coordination and each $Zr^{4+}$ ion is connected to four intra-cluster oxygen atoms, and four distinct carboxylate linkers (Figs. 1a, 1b).[22] The presence of such high coordinate inorganic centers is thought to confer a large shear modulus upon the framework, though rapid amorphization upon ball-milling is still observed.[18]

The two related MOFs, **MIL-140B** $[ZrO(O_2C-C_{10}H_6-CO_2)]$ and **MIL-140C** $[ZrO(O_2C-C_{12}H_8-CO_2)]$, also collapse relatively quickly upon ball-milling. Respectively crystallizing in the space groups $Cc$ and $C2/c$, each contains seven-coordinate $Zr^{4+}$ ions, which are connected to four 2,6-napthalene dicarboxylate (**ndc**) and 4,4'-biphenyldicarboxylate (**bpdc**) ligands respectively. Purely inorganic ZrO chains line the *c*-axis of the cells and delimit triangular channels (Fig. 1c), which are larger in **MIL-140C** due to the larger **bpdc** organic linker used (Fig. 1e). Whilst π-stacking along the *c*-axis between all **bpdc** ligands (i.e. those bisecting the lozenge-shaped channels) is observed in **MIL-140C**, only 50% of the corresponding **ndc** ligands in **MIL-140B** participate in this stabilization (Fig. 1d, 1f). [23]

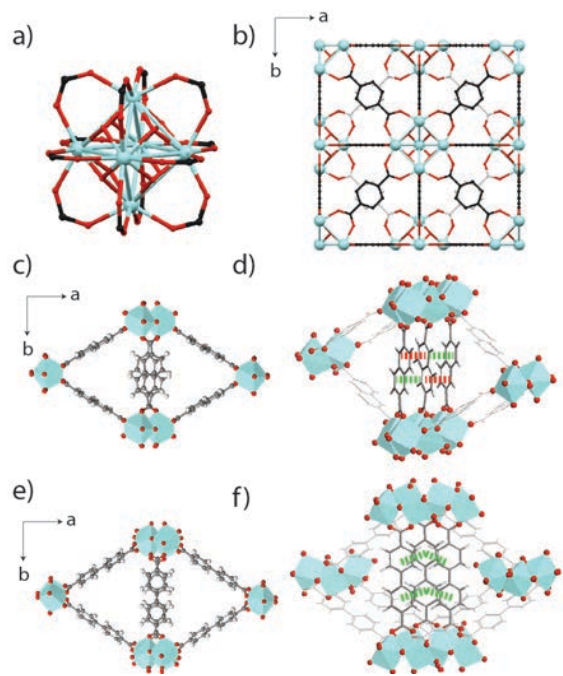

**Figure 1**. (a) The $Zr_6O_4(OH)_4$ inorganic node present in **UiO-66**. (b) Unit cell of **UiO-66** viewed along a cubic axis. Carbon atoms behind those shown in the foreground are coloured off-white for clarity. (c) The **MIL-140B** crystal structure proposed in ref [23], viewed along the *c*-axis, along with (d) π -stacking distances along the *c*-axis between the **ndc** ligands. Red – 4.46 Å, green – 3.63 Å. (e) The MIL-140C crystal structure proposed in ref [23], with (f) corresponding π -stacking distances (green – 3.94 Å) along the *c*-axis between the **bpdc** linkers. Zr atoms are depicted as light blue polyhedra, O is red, C is gray and H is white.

Connections between the mechanical properties of MOFs and their defect content are of interest. **UiO-66** has previously been demonstrated to be prone to linker vacancies,[22, 24] the extent of which can be tuned to yield differential adsorption and thermo-mechanical properties.[25, 26] Specifically, recent work has pointed to the direct coordination of $H_2O$ and solvent molecules to $Zr^{4+}$ nodes in defective UiO-type structures.[27, 28] Differences between simulated and experimental cell parameters of solvothermally synthesized MIL-140B point towards the presence of defects in this framework, though microwave based syntheses are recorded as producing a defect free material.[29, 30] Potential links between defects and amorphization mechanisms are therefore of much interest, alongside the structure and properties of the resultant amorphous products.

We use infrared spectroscopy and pair distribution function (PDF) measurements to structurally characterize the products of the ball-milling induced collapse of the rigid **UiO-66**, **MIL-140B** and **MIL-140C** frameworks. Density functional theory (DFT) calculations reveal the presence of defects in **MIL-140B**. We build defective models, and alongside $^{13}$C solid state magic angle spinning (MAS) nuclear magnetic resonance (NMR), use these to suggest the likely nature of the defects in the framework. Overall, we find a quantifiable partial breakage of metal-ligand bonding in each case, though, perhaps surprisingly, the backbone ZrO chains of the **MIL-140** frameworks are also damaged in the process. The [$Zr_6O_4(OH)_4$] clusters of **UiO-66** however remain intact.

## Results and Discussion

*Powder X-ray Diffraction and Infrared Spectroscopy*

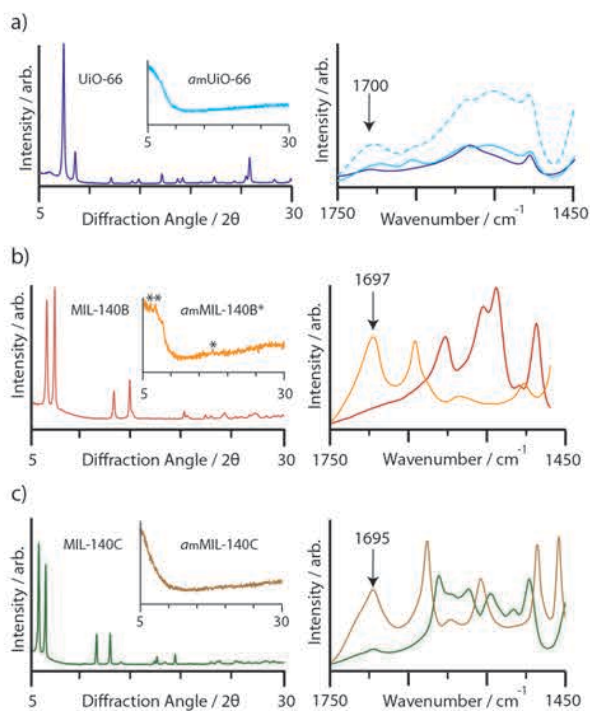

**Figure 2**. CuKα₁ powder X-ray diffraction patterns, and IR spectra, of crystalline and (inset) amorphous samples (left hand plots), complete with IR spectra of the region 1450 – 1750 cm-1 (right hand plots). (a) **UiO-66**: dark blue, $a_m$**UiO-66**: light blue, $a_m$**UiO-66** after 20 minutes ball milling: broken light blue, (b) **MIL-140B**: red, $a_m$**MIL-140B***: orange, (c) MIL-140C: green, $a_m$**MIL-140C**: brown.

Consistent with prior work, ball-milling of crystalline samples of **UiO-66** (10 mins), **MIL-140B** (15 mins) and **MIL-140C** (20 mins) resulted in the disappearance of Bragg peaks from the powder X-ray diffraction patterns (Fig. 2, SI-1). We term the products $a_m$**UiO-66** and $a_m$**MIL-140C**, where the subscript m refers to amorphization by mechanochemical means. An asterisk denotes the retention of some Bragg peaks in the diffraction pattern of **MIL-140B** after milling ($a_m$**MIL-140B***, Fig. 2b inset). Infrared spectroscopy revealed the emergence of a band centered at ca.1700 cm-1 upon amorphization of the **MIL-140** samples, which is assigned to the uncoordinated carbonyl stretching frequency (Fig. 2). An increase in intensity of this band was also noted upon collapse of **UiO-66**, where uncoordinated **bdc** ligands within the pores result in a small feature at 1700 cm-1 in the crystalline sample.[31]

*Nuclear Magnetic Resonance Spectroscopy*

The qualitative observation of partial destruction of metal-ligand bonding motivated us to perform solid-state $^{13}$C MAS NMR on the compounds. The experimental spectrum of **UiO-66** (Fig. 3a) contains three peaks. Based on prior work,[31] resonances at 128 ppm and 137 ppm are assigned to the two types of carbon on the **bdc** aromatic ring, whilst the carboxylate-binding group gives rise to the signal at 170 ppm. The $^{13}$C NMR spectrum of **MIL-140B** (Fig. 3b) is significantly more complicated, containing multiple signals in the 120 ppm – 140 ppm region and two distinct resonances at 173.5 ppm and 175 ppm. A further increase in complexity is noted in the $^{13}$C MAS NMR experimental spectrum of **MIL-140C** (Fig. 3c), which now contains three resonances in the region 170-175 ppm.

Upon amorphization, significant peak broadening of all signals is observed in the experimental spectra, though an indication of retention of the **bdc** ligand in $a_m$**UiO-66** is given by the two identifiable signals which make up the broad peak at low chemical shift (Fig. 3a). Coalescence of the signals belonging to non-carboxylate carbon atoms in $a_m$**MIL-140B*** is observed (Fig. 3b), while three distinct features in the 120-145 ppm $^{13}$C spectrum of $a_m$**MIL-140C** remain apparent (Fig. 3c). Loss of chemically distinct ligand environments in the **MIL-140** frameworks was confirmed by coalescence of carboxylate peaks at ca. 170 ppm in each case.

A small additional feature at ca. 182.5 ppm appears upon amorphization for all three samples (Fig. 3d). The higher chemical shift of this peak is consistent with loss of carboxylate coordination to the $Zr^{4+}$ ions upon structural collapse. Integration traces of the carboxylate signals in the amorphous samples yielded the approximate ratio of this emergent peak to the main peak. In $a_m$**UiO-66**, the new feature was found to account for ca. 6.8 % of the total intensity from the carboxylate carbon, suggesting a low degree of Zr-OOC bond breaking. This is raised to ca. 13.2 % and 11.0 % respectively in **MIL-140B** and **MIL-140C**, implying a higher degree of coordinate bond breaking in these samples upon amorphization.

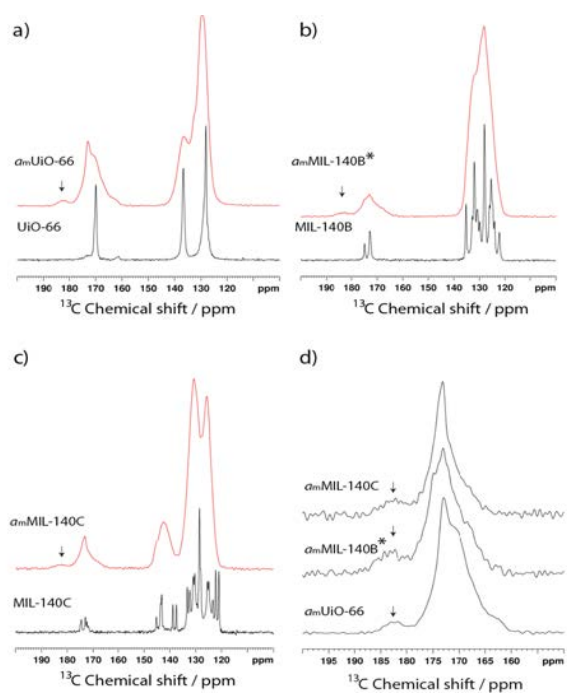

**Figure 3**. $^{13}$C MAS NMR spectra of crystalline (black) and amorphous (red) samples of (a) **UiO-66**, (b) **MIL-140B** and (c) **MIL-140C**. (d) Magnified view of the 150-200 ppm region of the amorphous samples. Acquisition parameter details are given in the SI.

*Non-defective modeling of MIL-140B*

In order to understand and better assign individual features in the experimental $^{13}$C NMR spectra of **MIL-140B** and **MIL-140C**, DFT calculations were performed using their previously reported crystal structures,[23] which were derived using a computationally assisted strategy from the crystal structure of **MIL-140A** (the bdc-based isoreticular structure); the only one solved by Rietveld refinement of X-ray powder diffraction data.[23] Simulated cell parameters and volumes for **MIL-140B** and **MIL-140C** were obtained in the present work upon fully relaxing coordinates and cell parameters, using four different DFT approaches; one with no dispersion correction (PBE), two with dispersion correction using the semi-empirical vdW method by means of D2 and D3 corrections (PBE-D2[32] and PBE-D3[33]), and a fourth one employing one of the non-local van der Waals functionals (optB88-vdW).[34] We first discuss attempts at structural optimization of alternative, non-defective **MIL-140B** structures, before considering defective **MIL-140B** models and their relative energetics, and then their respective calculated $^{13}$C NMR responses.

**Table 1**. Cell parameters of experimental and optimized crystal structures of **MIL-140B**, **MIL-140B(r)** and **MIL-140C**. **MIL-140B(r)** refers to a defective model imposing a 180° rotation around the *b*-axis to **ndc** linkers (see text for details). PBE-D2, PBE-D3, optB88-vdW represent dispersion corrected schemes. PBE-D3//exp entry refers to a geometry optimization of **MIL-140B** at the experimental lattice parameters.

| MIL-140B     | $a$ (Å) | $b$ (Å) | $c$ (Å) | $\alpha$(°) | $\beta$(°) | $\gamma$(°) | $V$(Å$^3$) | E (eV) |
|--------------|---------|---------|---------|-------------|------------|-------------|------------|--------|
| Exp. Ref 22  | 26.71   | 13.30   | 7.79    | 90          | 92.56      | 90          | 2763.2     |        |
| PBE          | 28.05   | 13.49   | 8.02    | 90          | 94.26      | 90          | 3024.7     |        |
| PBE-D2       | 28.05   | 13.46   | 7.87    | 90          | 93.40      | 90          | 2964.7     |        |
| PBE-D3       | 27.95   | 13.46   | 7.91    | 90          | 93.34      | 90          | 2970.7     | 0      |
| PBE-D3//exp  | 26.71   | 13.30   | 7.79    | 90          | 92.56      | 90          | 2763.2     | +1.57  |
| optB88-vdW   | 28.00   | 13.44   | 7.89    | 90          | 93.82      | 90          | 2963.3     | 0      |
| **MIL-140B(r)** |      |         |         |             |            |             |            |        |
| PBE-D2       | 27.65   | 13.41   | 7.84    | 89.32       | 90.12      | 88.86       | 2906.1     |        |
| PBE-D3       | 27.56   | 13.41   | 7.88    | 89.50       | 90.34      | 88.81       | 2909.8     | -0.38  |
| optB88-vdW   | 27.28   | 13.38   | 7.85    | 91.06       | 88.66      | 88.16       | 2864.8     | -0.47  |
| **MIL-140C** | $a$ (Å) | $b$ (Å) | $c$ (Å) | $\alpha$(°) | $\beta$(°) | $\gamma$(°) | $V$(Å$^3$) |        |
| Exp. Ref 22  | 31.03   | 15.51   | 7.82    | 90          | 93.26      | 90          | 3756.6     |        |
| PBE          | 31.52   | 15.63   | 7.97    | 90          | 87.06      | 90          | 3918.3     |        |
| PBE-D2       | 31.29   | 15.58   | 7.78    | 90          | 89.67      | 90          | 3794.7     |        |
| PBE-D3       | 31.25   | 15.59   | 7.82    | 90          | 89.32      | 90          | 3810.9     |        |
| optB88-vdW   | 31.16   | 15.59   | 7.78    | 90          | 89.90      | 90          | 3778.8     |        |

Upon comparison of DFT results with the experimental values for **MIL-140B** (Table 1), all employed DFT methods converge systematically to a larger cell volume than the experimental one (7.5 % error). This mainly emanates from the recurrent overestimation of the lattice constant *a* found at ~28 Å (4.8 % error), while *b* and *c* are predicted with excellent accuracy (within 1.5 % error). This cell expansion occurs when dispersion corrections are omitted (PBE entry), and is surprisingly maintained when they are taken into account, a result which is not sensitive to the methodology by which the dispersion is calculated (PBE-D2, -D3 and optB88-vdW entries). A further geometry optimization of **MIL-140B**, constraining its cell parameters to the experimentally determined values (PBE-D3//exp entry), reveals it is much less stable by 1.57 eV than its relaxed **MIL-140B** counterpart (PBE-D3 entry), pointing towards an inconsistency between the structural model used and the experimentally determined cell parameters. Expectedly, the simulated $^{13}$C spectrum of **MIL-140B** is in very poor agreement with the experimental one, in both regions of carboxylate and aromatic carbons (Fig. S1a), supporting a hypothesis of a different, or defective MIL-140B structure not captured in the initial model of ref [22].

The situation is markedly different with **MIL-140C**, with excellent predictions from DFT structure optimizations (Table 1 and Fig. S2). While dispersion correction-free DFT calculations slightly overestimate the *a* cell parameter (1.6% error) and cell volume, *V* (4.3% error), all dispersion-corrected DFT calculations yield excellent agreement when compared to experiment, particularly the optB88-vdW entry with errors reduced to 0.4 % and 0.6% for *a* and *V*, respectively. As in all **MIL-140B** calculations, the *c* parameter is systematically very well predicted. This reflects the structural rigidity associated with the inorganic subunit of the zirconium oxide chains along that direction. Taking the dispersion effects into account improves the simulation of the **MIL-140C** crystal structure and highlights the importance of the π-stacking stabilizing interactions along the *c*-axis, which take place with the involvement of all **bpdc** ligands at regular center-to-center distances of 3.9 Å (Fig. 1f). The simulated NMR spectra provide significantly better agreement with observed NMR shifts, supporting a consistent initial structural model (Fig. S2).

The failure of DFT calculations to predict the correct cell volume and lattice parameters for **MIL-140B** reveals that its stabilization at the smaller observed cell parameters emanates from additional structural features. This together with NMR discrepancies prompted us to an in-depth investigation of the structure and defects in **MIL-140B**, as a case study.

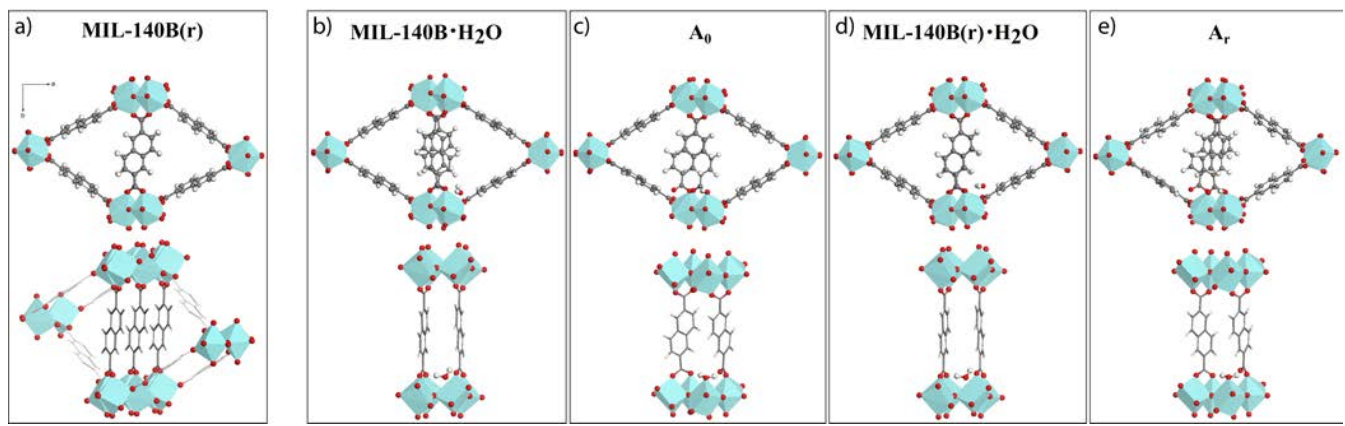

**Figure 4**. Crystal structure of **MIL-140B(r)**, formed by rotating 50% of the ndc linkers which lie along the *a*-axis, around the *b*-axis by 180°. Structures of **MIL-140B** contain one $H_2O$ molecule per unit cell viewed along the *c*-axis. Models called **MIL-140B•H2O** and **MIL-140B(r)•H$_2$O** contain an uncoordinated water molecule, while **A$_0$** and **A$_r$**, are defective models and are built with a $H_2O$ molecule coordinates directly to a Zr-center. The latter displaces the linker which becomes monodentate on one its carboxylate end. Zr atoms are depicted as light blue polyhedra, O is red, C is gray and H is white.

## *Enhanced π Stacking in MIL-140B*

A first variant of the reported **MIL-140B** structure, hereby referred to as **MIL-140B(r)**, where (r) indicates the partial rotation of linkers, was therefore constructed by imposing a 180° rotation around the *b*-axis to the **ndc** linkers not participating in the π-stacking (Fig. 4a). The resultant **MIL-140B(r)** model is hence one in which a full π-stacking of **ndc** linkers along the *c*-axis occurs. As a result, the fully relaxed **MIL-140B(r)** model is 0.38 eV lower in energy than the parent **MIL-140B**, compared at the same level of theory (Table 1, PBE-D3 entry). This stabilization is assigned mainly to the now enhanced π-stacking along *c*. In addition, the sole rotation of these **ndc** linkers induces a large contraction along the *a*-axis, of 0.4-0.7 Å depending on the level of theory, with an error reduced to 2-3%.

*Addition of uncoordinated H₂O*

While being closer to experimental findings, **MIL-140B(r)** still exhibits overestimated cell parameters. Given recent reports of the defects in UiO-materials being caused by $H_2O$ or solvent molecules, [27, 35] we thus further constructed several non-defective structures, where low concentrations of water are introduced as adsorbed molecules. Table 2 summarizes the optimized cell parameters and their relative energies obtained at the PBE-D3 level of theory.

Two hydrated models were derived from **MIL-140B** and **MIL-140B(r)**, respectively (labeled "•$H_2O$"), by adding one adsorbed water molecule per cell (Fig. 4b, 4d). Upon full relaxation, the water molecule preferentially forms hydrogen bonds with two carboxyl oxygen atoms (1.99 and 1.95 Å) of two neighboring **ndc** linkers stacked along *c*, with minor change in cell volumes. The rotated hydrated model was found to be of lower energy (by 0.35 eV) than the non-rotated one, which is ascribed entirely due to the enhanced π-stacking in **MIL-140B(r)** (see Table 1). Addition of a second uncoordinated water to both models (labeled •$2H_2O$), to a site at the opposite end of the organic linker to the first molecule, did not result in any cell modification nor further stabilization.

## *Defect Modelling*

*$H_2O$ coordination to Zr combined with linker displacement*

In further models, $H_2O$ was allowed to coordinate to $Zr^{4+}$ ions, forming defective **MIL-140B** structures. Four possible defect incorporation pathways, (a)-(d) were considered (Figure 5), respectively leading to models **A-D** (Fig. S3). Models arising from the original **MIL-140B** structure are given the subscript '0', and those arising from **MIL-140B(r)**, are given the subscript 'r', e.g. **A₀** and **A_r** respectively for pathway 'a'.

**Table 2**. Optimized cell parameters (PBE-D3 level of theory) of **MIL-140B** defective variants containing different amount of adsorbed or coordinated water molecules per unit cell. Energies are given in eV with respect to the structures with adsorbed (not coordinated) water molecules. 'r' refers to models constructed with rotated linkers for π-stacking enhancement while "0" refers to models constructed from the parent **MIL-140B**. A, B, C and D refer to the various defect pathways (a, b, c, d) in Figure 5. 'e' refers to models with extended coordination of water molecules along the *a*-axis.

| Uncoordinated Water | $a$ (Å) | $b$ (Å) | $c$ (Å) | $\alpha$(°) | $\beta$(°) | $\gamma$(°) | $V$(Å$^3$) | E (eV) |
|---|---|---|---|---|---|---|---|---|
| MIL-140B•H$_2$O | 27.84 | 13.47 | 7.89 | 90.57 | 92.79 | 90.02 | 2957.5 | 0 |
| MIL-140B•2H$_2$O | 27.82 | 13.48 | 7.90 | 89.96 | 93.01 | 90.54 | 2955.5 | 0 |
| MIL-140B(r)•H$_2$O | 27.52 | 13.42 | 7.87 | 89.49 | 90.15 | 88.94 | 2905.4 | -0.35 |
| MIL-140B(r)•2H$_2$O | 27.51 | 13.44 | 7.86 | 89.26 | 90.24 | 89.15 | 2905.4 | -0.37 |
| | | | | | | | | |
| **Coordinated Water** | | | | | | | | |
| **A$_r$** | 27.28 | 13.42 | 7.86 | 90.07 | 88.28 | 89.20 | 2875.1 | +0.02 |
| **A$_{re}$** | 26.63 | 13.45 | 7.83 | 90.36 | 84.44 | 89.91 | 2789.4 | +0.40 |
| **A$_0$** | 27.79 | 13.46 | 7.87 | 90.34 | 92.13 | 90.52 | 2944.2 | +0.21 |
| **A$_{0e}$** | 27.56 | 13.49 | 7.85 | 90.79 | 90.12 | 91.10 | 2918.8 | +0.37 |
| **B$_r$** | 27.40 | 13.41 | 7.85 | 90.67 | 89.54 | 87.65 | 2882.5 | +0.26 |
| **B$_0$** | 27.69 | 13.47 | 7.85 | 89.43 | 92.26 | 89.51 | 2925.6 | +0.54 |
| **C$_0$** | 27.77 | 13.46 | 7.85 | 90.18 | 93.30 | 89.41 | 2930.2 | +0.82 |
| **D$_r$** | 27.10 | 13.45 | 7.86 | 89.45 | 87.19 | 89.77 | 2861.3 | +0.37 |

In pathway 'a', one end of an **ndc** linker was displaced into a monodentate state, allowing coordination of one H$_2$O molecule to the newly created defect site. Upon relaxation, the chemisorbed water remained bound to Zr and established hydrogen bonds with its two surrounding carboxyl oxygen atoms (1.80 Å). Interestingly, model **A$_r$** is isoenergetic to the relative ground state **MIL-140B•H$_2$O** structure (0.02 eV): the energy penalty for breaking Zr-O(linker) bond (~0.3 eV) and altering the π-stacking is compensated by the coordination of the water molecule to Zr. In addition, unlike the (non-coordinating) adsorption of water, the chemisorption of water on the Zr-center in **A$_r$** leads to a significant cell contraction along *a* (up to~0.3 Å) when compared to the water-free model, making $A_r$ a promising defective model candidate.

Extension of this defect with a second coordinated $H_2O$ molecule along the *a*-axis (model $A_{re}$) induces a relatively low energetic penalty (+0.40 eV). More importantly, this results in a further cell contraction of this defective **MIL-140B** model of the *a* lattice constant (*a*= 26.63 Å), in excellent agreement with the experimentally determined value for the **MIL-140B** structure of 26.71 Å.

Identical treatment of **MIL-140B•$H_2O$**, the non-rotated system, gave rise to models $A_0$ and $A_{0e}$, which are only 0.21 eV and 0.37 eV higher in energy than the reference ground state of **MIL-140B** with uncoordinated water molecule, respectively, with however again poor agreement in lattice parameters.

Turning to defect B, we considered translation of the entire **ndc** linker induced by the chemisorption of two water molecules on the opposite Zr-centers (pathway 'b' in Figure 5). As a result, the linker becomes monodentate on one side and bidentate on the other. The resultant models, $B_r$ and $B_0$, are +0.26 and +0.54 eV higher in energy than the reference model, respectively.

A third type of defect yielded model $C_0$ (pathway 'c', in Figure 5). Whilst the defect resembles that in pathway 'b', the linker binds to two neighboring Zr-centers along the *c* direction. This structure was deemed unfavourable by +0.82 eV with respect to the reference defect-free model **MIL-140B•2H2O**.

A fourth defect structure, modeled by following pathway 'd' in Figure 5, yields $D_r$. In this structure both ends of an **ndc** linker were displaced into a monodentate position, allowing coordination of a $H_2O$ molecule to each of the two newly created defective Zr sites. It can form with an energy cost of 0.37 eV and shows significant shrinking of the lattice vector *a* to 27.10 Å.

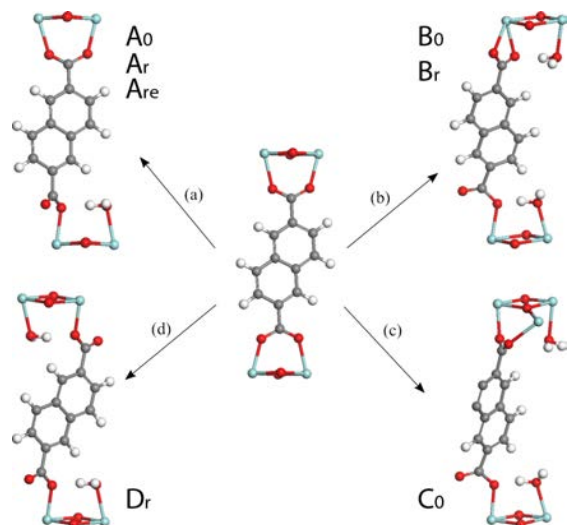

**Figure 5**. Formation of different types of defective structures upon direct binding of water on a Zr-center. The names of the defective models derived from each pathway are included for clarity. Zr – light blue, O – red, C – gray, H – white.

Overall, this series of calculations tend to suggest that the formation of defects in **MIL-140B** may occur through the coordination of water molecules to Zr centers together with the displacement of the linker and that such defects are associated with relatively low energy penalties. They also indicate that defect pathways allowing the enhancement of π-stacking of **ndc** linkers along *c* direction might be favored, exemplified in 'r' type models.

*Comparison of predicted 13C NMR shifts from defective models*

$^{13}$C NMR chemical shift calculations were performed on our defective **MIL-140B** variants in order to evaluate the pertinence of these models with respect to available experimental data. Calculation of the $^{13}$C NMR chemical shifts for the **MIL-140B(r)** model yielded significantly better agreement with the experimentally observed shifts than those attained directly from **MIL-140B** (Fig. S1b). This is most notable in the increased complexity of the splitting pattern in the aromatic carbon region that covers more consistently the whole range of observed peaks. The concordance of such splitting between the experimental data and this **MIL-140B(r)** model suggests that rotation of **ndc** linkers might indeed occur so as to favor π-stacking locally.

This improved agreement in the 120-140 ppm aromatic region is retained in hydrated models, i.e. $\mathbf{A_{re}}$, $\mathbf{B_r}$ and $\mathbf{D_r}$, in which H$_2$O is introduced directly to the Zr coordination sphere. The region of the NMR spectra attributed to carboxylate shifts (170-190 ppm) of $\mathbf{A_{re}}$, $\mathbf{B_r}$ and $\mathbf{D_r}$, however, shows a lower level of agreement with the experimental spectra of **MIL-140B** with a large dispersion of the calculated values compared to the experimental ones.

Whilst possible, defect-incorporating models ($\mathbf{A_{re}}$, $\mathbf{B_r}$ and $\mathbf{D_r}$), for the reported **MIL-140B** structure have been suggested, it is equally interesting that some of the features in the predicted NMR shifts agree well with those of $a_m$**MIL-140B**. The range of chemical shifts across the spectra observed shows excellent agreement with those for the amorphized **MIL-140B** sample. The predicted $^{13}$C NMR spectrum of model $\mathbf{B_r}$ however, is the only one to exhibit a peak in the region 185 ppm – 190 ppm, which is identified to be a fingerprint of one carboxylate end being bound in a bidentate fashion to Zr$^{4+}$. Since this is closely reminiscent of the feature which appears upon amorphization, it provides some evidence as to the introduction of defects upon ball-milling of the structure.

*Pair Distribution Function Studies*

Analysis of the pair distribution function, or the weighted histogram of atom-atom distances,[36] of amorphous inorganic zeolites[37] and MOFs,[12] has in the past yielded useful information on the chemical structure of complex amorphous systems. Room temperature total scattering data were therefore collected on crystalline and amorphous samples using synchotron radiation ($\lambda$ = 0.1722 Å, $Q_{max}$ = 22 Å$^{-1}$). The resultant structure factors $S(Q)$ (Figs. S4-S6) of $a_m$**UiO-66** and $a_m$**MIL-140C** are devoid of Bragg peaks, confirming the amorphous sample nature. Expectedly, that of $a_m$**MIL-140B*** contains some features due to Bragg scattering. After suitable corrections using the GudrunX software,[38] the data were converted to the corresponding PDFs by Fourier Transform.[39] To facilitate assignment of features, partial PDFs were calculated using the PDFgui software (Figs. S7-S9).[40] All PDFs of crystalline species are, as expected, dominated by correlations involving Zr, due to the larger X-ray scattering cross section of Zr relative to C, O and H.

Due to the difficulties in obtaining reliable data below ~ 1 Å from X-ray total scattering instruments, the first peak in the PDF of crystalline **UiO-66** belonging to a physical atom-atom correlation appears at ~ 1.3 Å, which corresponds to C-C or C-O direct linkages (A, Fig. 6a). Other features below 6 Å (B-F) are assigned to various Zr-Zr and Zr-O inter-cluster separations. At longer distances, overlapping contributions from similarly spaced Zr-C atom pairs results in peak broadening and renders precise assignment of the features at ~ 6.5 Å and ~ 8.6 Å challenging. The large relative scattering cross-section of Zr results in sharp features above 10 Å (G-K) in the PDF of **UiO-66**, which can be ascribed to Zr atom pairs joined through a **bdc** linker (Fig. 6b).

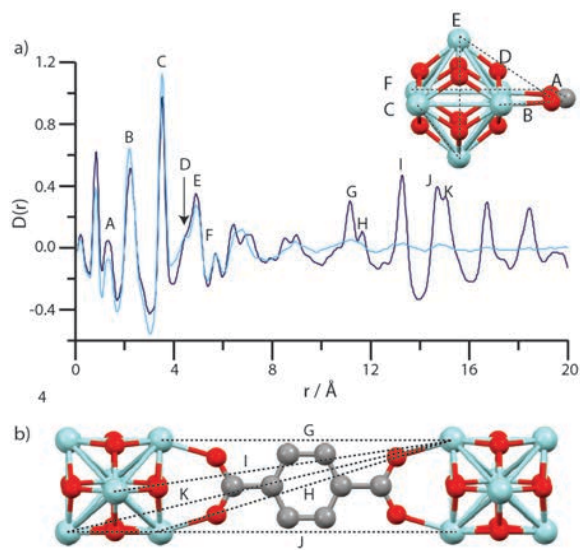

**Figure 6**. (a) PDF data for **UiO-66** (dark blue) and $a_m$**UiO-66** (light blue). A-F labels of peaks below 8 Å correspond to the indicated correlations in the $Zr_6O_4(OH)_4$ cluster (Inset). (b) Two $Zr_6O_4(OH)_4$ units linked by a **bdc** ligand, and some of the significant distances corresponding to the longer $r$ features (G-K) in (a). Zr – light blue, O – red, C – gray, H – omitted.

Below 6 Å, the PDF of $a_m$UiO-66 is very similar to UiO-66, with peaks A-F appearing invariant in position between the two and thus confirming the presence of the $Zr_6O_4(OH)_4$ structural building unit in $a_m$UiO-66. Coalescence of the double peaks centered at ~ 6.5 Å and ~ 8.6 Å occurs upon amorphization, yielding two broad features in the PDF of $a_m$UiO-66. The reduction in intensity and severe broadening of these peaks, and those labeled G-K, in $a_m$UiO-66 is indicative of the loss in long range-order of the framework. The presence however of some intensity does suggest at least the partial retention of the structural linkage, though substantial changes to Zr-Zr correlations would be expected upon breaking if only one Zr-O bond were to break. The observable extent of order in $a_m$UiO-66 may extend beyond 15.9 Å, though this is the last distance at which correlations can be unambiguously assigned to Zr-Zr atoms separated by the organic linker.

The PDF of MIL-140B (Fig. 7a) contains the expected peaks (L-N) belonging to nearest neighbour C-C/C-O, Zr-O and Zr-Zr distances at ca. 1.3 Å, 2 Å and 3.3 Å, whilst at further distances below 8 Å (O-R), non nearest neighbor correlations of Zr-Zr and Zr-O atoms also elicit sharp features. These corresponding chemical distances relating to L-R are indicated in the inset of Figure 7a, and all relate to the inorganic ZrO backbone chain. Sharp features associated with crystalline order continue to be observed in the PDF of MIL-140B to distances up to 20 Å, as expected. Key features (S-U) at these longer $r$ values agree well with Zr-Zr distances separated by the 2,7-ndc linker (Fig. 7b).

Features L and M are located in the same position in the PDF of $a_m$MIL-140B* (Fig. 7a), though N, which corresponds to the nearest Zr-Zr distance in the inorganic backbone appears to lengthen from 3.3 Å to 3.5 Å upon amorphization. Another noticeable change in the PDF of $a_m$MIL-140B* is the loss in intensity of peaks O-Q, which, correspond to non-nearest neighbour Zr and O distances in the ZrO backbone chains. Somewhat surprisingly, the peak at 7.8 Å, labeled R, disappears entirely on going from MIL-140B to $a_m$MIL-140B*, suggesting the rigid ZrO inorganic chain no longer remains intact when longer Zr-Zr correlations are considered. Indeed, changes are apparent at even shorter Zr-Zr length scales, with substantial deterioration of the intensity of the Zr-O-Zr (O) peak intensity at 4.5 Å, which suggests that some Zr-O bonds have been broken during the ball-milling process, causing the chain to twist. Given this partial destruction of the continuous ZrO backbone, it is surprising that pair correlations relating to Zr atoms separated by the organic linker (S, T and U) persist in $a_m$**MIL-140B***, with some degree of correlated order extending to ca 15.9 Å. Above this distance, the PDF of $a_m$**MIL-140B*** is featureless, in sharp contrast to its crystalline counterpart.

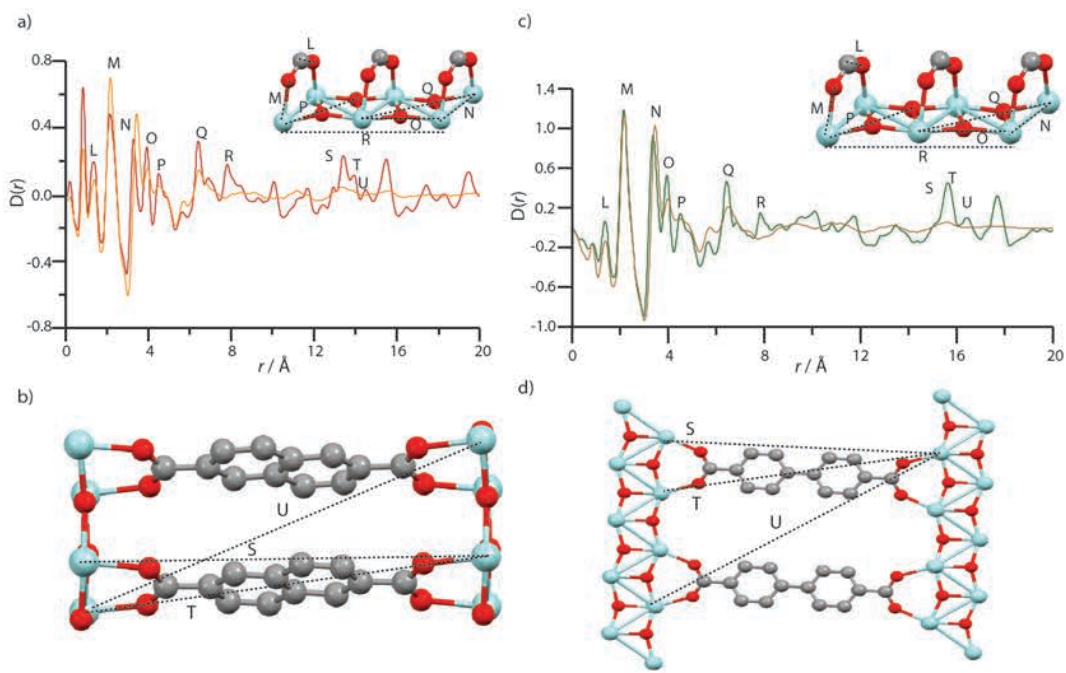

**Figure 7**. (a) PDF data for **MIL-140B** and $a_m$**MIL-140B***. Labels of peaks below 8 Å correspond to the indicated correlations in the ZrO inorganic chains (Inset). (b) Two ZrO chains linked by 2,6-**ndc**, and the distances corresponding to the longer *r* features in (a). (c) PDF data for **MIL-140C** and $a_m$**MIL-140C**. Labels of peaks below 8 Å correspond to the indicated correlations in the ZrO inorganic chains, as in **MIL-140B** (Inset). (d) Two ZrO chains linked by bpdc, and the distances corresponding to the longer *r* features in (a). Zr – light blue, O – red, C – gray, H – omitted.

PAGE 2

The near-identical nature of the **MIL-140C** framework, aside from the slightly longer organic linker, led to the labeling of corresponding distances in the PDFs with the same notation as that used in **the MIL-140B** PDF. Similar differences between amorphous and crystalline frameworks are observed in the PDFs of **MIL-140C** and $a_m$**MIL-140C** (Fig. 7c, d). In particular, the same lengthening of nearest Zr-Zr distances (N) upon amorphization is witnessed. In addition, very similar reductions in the intensity of the peak belonging to nearest neighbor Zr-O correlations at 4.5 Å (O) are also observed whilst there is very little intensity in further Zr-O correlations (P-R).

This observation provides confidence that the ZrO backbone shared by both **MIL-140B** and **MIL-140C**, undergoes significant distortion during ball-milling, though some order persists in the PDF of $a_m$**MIL-140C**, to 17.6 Å (S, T and U). These distances, as in $a_m$**MIL-140B***, correspond to two Zr atoms separated by the larger **bpdc** organic linker (Fig. 7d).

## Conclusions

The results further contribute to the area of non-crystalline metal-organic frameworks and provide insight into the relationship between defects and amorphization. We have identified possible modes of defect incorporation into the structure of **MIL-140B**; Defects investigated include coordinated water to Zr centers combined with linker displacement. Such defective models provide significantly better agreement with the experimentally observed structural cell parameters than the defect free ones. DFT calculations of the chemical shifts of some models also show similar features to those witnessed experimentally upon amorphization of **MIL-140B**. We anticipate that the occurrence of such defects might be applicable to other MOFs, given the prevalence of carboxylate linkers in this family.

In addition to solid state NMR, we have used pair distribution function analysis to show that the amorphization of the prototypical **UiO-66** framework proceeds via partial breakage of Zr-carboxylate bonds, though the rigid $Zr_6O_4(OH)_4$ inorganic building unit appears to remain intact. This is in stark contrast to **MIL-140B** and **MIL-140C**, where in addition to metal-carboxylate bond breaking, the inorganic backbone is also heavily damaged during the ball-milling process. This mechanism bears a large similarity to the changes seen in the ball-milling of zeolites, where M-O (M = Si, Al) bonds are observed to be broken,[41] whilst the inorganic distortions are to be expected given that pure Zirconia undergoes phase transitions and eventual amorphization when subjected to ball-milling.[42]

The different behavior of the two inorganic units is ascribed to the larger interconnected nature of the inorganic clusters in **UiO-66**, where 24 Zr-O bonds hold the unit together. This is in contrast to the ZrO chains of the **MIL-140** frameworks, where in a unit containing the same number of Zr ions, only 16 Zr-O bonds hold the chain together. Whilst structural defects will also play a role in determining mechanical stability, the research here may be combined with increasing metal-ligand bond strengths,[20] to yield 'strong' MOFs capable of resisting external stresses.

**Author Contributions:** The manuscript was written through contributions of all authors. TDB conceived the initial project.

**Corresponding Author**: tdb35@cam.ac.uk, caroline.mellot-draznieks@college-de-france.fr

**Acknowledgements:** T.D.B. acknowledges Trinity Hall (University of Cambridge) and Professor Anthony K. Cheetham for use of lab facilities. D.G.R. acknowledges the UK MRC for financial support. The authors acknowledge Diamond Light Source for the provision of synchrotron access to Beamline I15 (exp. EE9691) and Philip A. Chater and Andrew Cairns for assistance with data collection. T.K.T. and C.M.D. thank the French National Research Agency (ANR project: HOPFAME ANR-13-BS07-0002-01) and the Foundation de l'Orangerie for funding. The calculations have been performed using the HPC resources from GENCI (CINES/TGCC/IDRIS) through Grant (2015-097343 and -091461). B.B., B.V.d.V. and D.D.V. gratefully acknowledge the FWO for funding (aspirant grant).